# Complete structure of topographic maps in ephrin-A deficient mice.

Dmitry N. Tsigankov and Alexei A. Koulakov

*Cold Spring Harbor Laboratory, Cold Spring Harbor, NY 11724*

**Axons of retinal ganglion cells establish orderly projections to the superior colliculus of the midbrain. Axons of neighboring cells terminate proximally in the superior colliculus thus forming a topographically precise representation of the visual world. Coordinate axes are encoded in retina and in the target through graded expression of chemical labels. Mapping based on chemical labels alone does not yield required specificity of connections. Additional sharpening is provided by electric activity, which is correlated between neighboring axons. Here we propose a quantitative model which allows combining the effects of chemical labels and correlated activity in a single approach. Using this model we study a complete structure of two-dimensional topographic maps in mutant mice, in which one of the labels, encoding horizontal retinal coordinate (ephrin-A), is reduced/eliminated. We show that topographic maps in ephrin-A deficient mice display granular structure, with the regions of smooth mapping separated by linear discontinuities reminiscent of fractures observed in the maps of preferred orientation. We also demonstrate the presence of point singularities in topographic maps.**

Neural development leads to the establishment of precise pattern of connectivity in the nervous system. In many cases connectivity is organized into maps which serve to orderly represent various sensory inputs to the brain[1]. For example, two-dimensional surface of the retina is represented continuously in mammalian superior colliculus (SC). To implement this topographically precise mapping retinal ganglion cell (RGC) axons reach SC and terminate according to the coordinates of their origin in retina. Studies of topographic map development have focused on understanding the mechanisms underlying the geometrically precise sorting of RGC axons in SC[2].

Many factors contribute to the formation of precise topographic maps. Following the original chemoaffinity hypothesis of Sperry[3], a variety of chemical labels were discovered, which encode reciprocal coordinates in retina and SC[4-7]. Thus, the nasal-temporal (NT) axis in retina is encoded by graded expression of EphA receptor tyrosine kinases by RGC axons. The recipient rostral-caudal (RC) coordinate in SC is established by graded expression of ephrin-A (Fig. 1A), which can bind and activate EphA receptors and transmit to RGC axons information about their position in SC. A similar chemical marking system, involving an EphB/ephrin-B receptor/ligand pair, exists for the mapping of dorso-ventral (DV) axis of retina to the medial-lateral (ML) direction of SC (Fig. 1B). The two approximately perpendicular expression profiles appear to be in place to determine the correct termination sites by RGC axons[8-11].

The precision of axonal projections is further enhanced through mechanisms based on correlated neural activity[12]. Due to the presence of retinal waves during development electric activity is similar in RGC axons neighboring in retina[13-15]. Correlated activity therefore provides additional information about axonal geometric origin in retina and contributes to the precision of topographic projection[14].

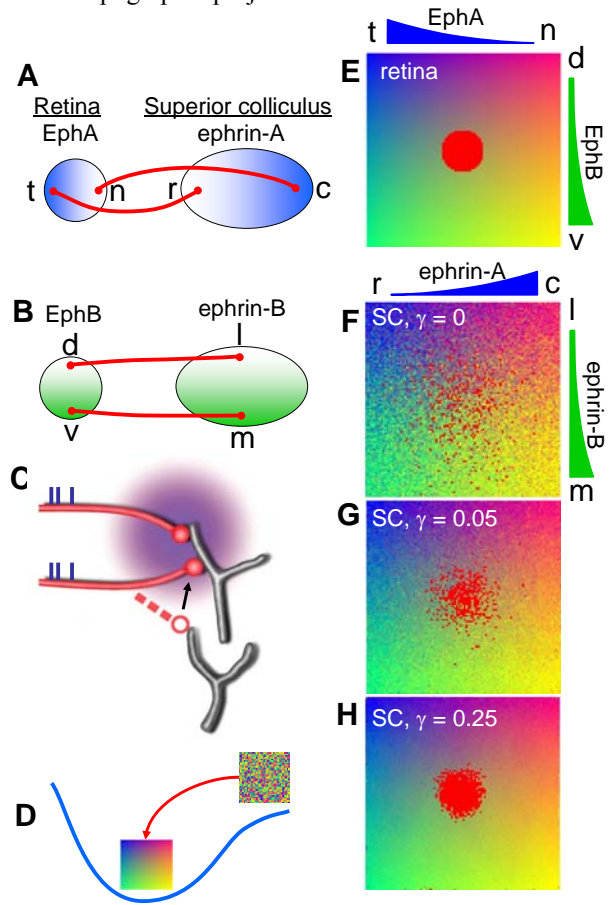

**Figure 1** Development of the topographic map in wild-type animals. (**A**) Ordering of axons along TM retinal axis is controlled by the EphA/ephrin-A receptor/ligand pair. (**B**) Ordering along DV axis is dependent upon EphB/ephrin-B pair. (**C**) Hebbian plasticity leads to an effective attraction of axons (red) with correlated activity (blue bars). The range of attraction (blue halo) is controlled by an average overlap between recipient dendritic arbors (gray). (**D**) Single axons follow the gradients of chemical labels[16], thus optimizing their affinity to the target. Here we assume that the entire system of axons representing retinocollicular mapping follows the gradient of generalized cohesive energy. This description is necessary, since axons compete with each other, which makes single-axon description inaccurate. (**E-H**) Activity sharpens retinocollicular projections by inducing attraction between axons neighboring in retina (red dots). Stronger activity leads to more precise connectivity (**G** to **H**). t, n, r, c, d, v, l, and m stands for temporal, nasal, rostral, caudal, dorsal, ventral, lateral, and medial correspondingly.

The layout of connectivity shown in Figs. 1A implies repulsion between EphA expressing axons and ephrin-A rich regions of SC[17]. If repulsive effect of ephrin-A exists, why do not all axons terminate in rostral SC, where the density of repellent is minimal (Fig. 1A)[2]? According to the competition hypothesis axons are distributed along the RC axis by mutually exclusive interactions, which limit the number of axons terminating in a fixed volume of the target[8, 11, 18-21]. Similarly, correlated gradients of EphB and ephrin-B along the ML direction (Fig. 1B) suggest affinity between RGC axons and rich in ephrin-B ligand areas[6], which is counterbalanced by either bifunctional action of ephrin-B ligands of axons[22] or axonal competition for space[23]. In both RC and ML cases competition between axons emerges as an important mechanism involved in the map formation.

In this study we combine three factors of retino-collicular map development, chemoaffinity, activity-dependent refinement, and competition, in a single quantitative model. We apply our model to maps in ephrin-A deficient mice[11], in which chemical signal is reduced along one of the directions. We reproduce experimental observations involving tracing of limited number of RGC axons from retina to SC. In addition, the use of computational model allows us to reconstruct complete connectivity between model retina and SC and fully visualize the topographic map in ephrin-A deficient mutants. We show that the topographic map under the influence of conflicting requirements of chemoaffinity and correlated activity carries features reminiscent of mappings in higher visual areas, such as preferred direction and orientation maps[24-26]. Our study helps to understand the logic of interactions between different factors contributing to topographic map formation and suggests further experimental tests.

**RESULTS**

We briefly introduce our model here to make the results clear. Our approach is based on optimization of cohesive energy between the system of RGC axons and their targets[27]. Cohesive energy describes both the energy of direct chemical binding in Eph/ephrin receptor/ligand pairs and the indirect effects of receptor activation and correlated activity. Because binding energy is expected to be minimal in the most stable axonal configuration, the cohesive energy is minimized in our approach (Fig. 1D). Minimum of the total cohesive energy also corresponds to the maximum of the total axonal affinity to their targets as was proposed by Sperry[3].

Cohesive energy includes both the energy associated with chemical binding and activation of Eph/ephrin receptor/ligand pairs and correlated-activity dependent contribution

$$E = E_{chem} + E_{act}. \quad (1)$$

Because activity block ($E_{act} = 0$) spares crude topographic ordering[2, 12, 14], presumably, due to remaining chemical labels (Ephs and ephrins), we choose an additive form of energy function. Thus, either chemoaffinity *or* correlated-activity (see below) is needed to establish topographic ordering.

The chemoaffinity contribution $E_{chem}$ results from binding and activation of Eph receptor by ephrin ligands. We first describe contribution from the EphA/ephrin-A pair. For each RGC axon the contribution to the chemical cohesive energy is given by the law of mass action $\Delta E_{chem} = \alpha [R_A][L_A]$, where $[R_A]$ is the level of available EphA receptor on this axon and $[L_A]$ is the level of available ephrin-A ligand at the termination site[27, 28]. To minimize the contribution to the cohesive energy axons move to the position with lower ligand level $[L_A]$, since the concentration of receptor $[R_A]$ is fixed for each axon. Minimization of $E_{chem}$ thus results in chemorepulsion between RGS axons and ephrin-A expressing regions in SC. To prevent all axons from terminating at the position with the lowest value of the ligand (rostral) we use competition between axons for space[8, 11, 18-21]. Such competitive interactions may arise due to limiting trophic factors in the target and/or direct axon-axon repulsion[11]. We employ the simplest model for competition in which each axon can occupy a fixed volume in SC. The problem of axonal sorting in SC becomes more intricate in the presence of competition. Since axons cannot all terminate in rostral SC, where the density of repellent is minimal, axons with high receptor levels (temporal) push axons with lower receptor levels (nasal) up the gradient of ligand into caudal SC. Thus, competitive interactions lead to the establishment of correct topographic ordering (Fig. 1A, F)[8, 11]. Similarly, in the perpendicular DV/LM direction (Fig. 1A), due to chemoattraction in the EphB/ephrin-B pair, RGC axons with high level of EphB receptor (ventral) terminate medially, in the high density of ephrin-B ligand, and push dorsal axons, which are attracted less strongly to ephrin-B, into lateral SC. We do not assume binding between EphA/B receptors and ephrin-B/A ligands respectively[5, 29-32] (see Methods for more detail).

The remaining part of equation (1) describes the correlated-activity dependent contribution, which we derive in Methods from the Hebbian learning rule. According to our calculations Hebbian plasticity leads to the attraction in SC between axons, which carry correlated electric activity, i.e. originate from neighboring points in retina. We illustrate the correlated activity dependent attraction in Fig. 1C. When two axons with correlated activity terminate on the same dendrite they drive postsynaptic cell effectively, leading to reinforcement of their connections through Hebbian mechanisms, according to which cells that fire together, wire together[33]. Alternatively, when two axons terminate on different dendrites their terminals are not reinforced by Hebbian mechanisms. Thus, Hebbian rule favors proximal positioning of coactive axons[34], which implies attraction. Such an attraction is described by two parameters in our model. The overall strength of attraction is characterized by coefficient $\gamma$ ($\gamma = 0$ implies no attraction, Fig. 1F), while the average range of axonal interactions (blue halo in Fig. 1C) is determined by the dimensions of recipient dendritic tree in SC.

From the standpoint of topographic projection, affinity of correlated axons to each other implies more precise mapping. Indeed, in Fig. 1G and H axonal terminals originating from close points in retina (red dots) are attracted to each other in SC, which leads to their more compact positioning and, therefore, more precise map. Our model captures topographic map refinement dependent on correlated neural activity.

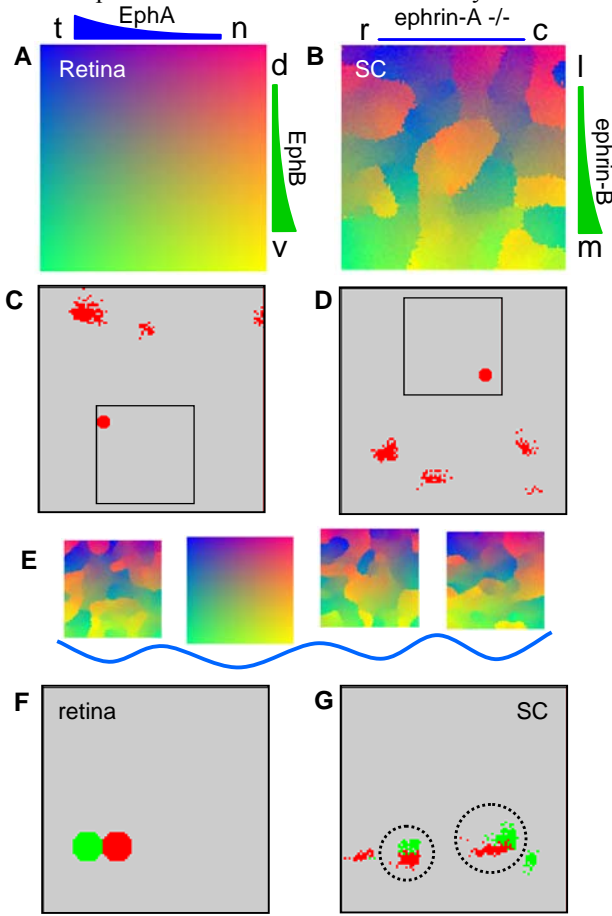

**Figure 2** Mapping in complete ephrin-A knockouts is discontinuous. Map from retina (**A**) to SC (**B**) displays granules of smooth mapping separated by linear discontinuities. (**C**) Tracing of connections originating from a circular region in temporal retina (inset) produces patchy clusters in SC similarly to Ref. [11]. Comparison to (**B**) shows that these clusters correspond to three granules originating from a blue region in the retina. (**D**) Nasal connections terminate in four disconnected clusters. These clusters correspond to four yellow granules in (**B**). Loss of DV precision in the mapping is apparent. (**E**) Due to lack of chemical labels in TN direction global minimum of the cohesive energy, corresponding to the ordered map, is obscured by local minima. Mapping terminates in one of this minima leading to the lack of global order. Three different disordered maps result from restarts of minimizing procedure from random initial conditions. (**F**, **G**) Paired retinal injections lead to paired clusters in SC (circles) reflecting local topography.

The final result of our model is the complete topographic map from retina and SC. To obtain the map we stochastically minimize total cohesive energy (1) with respect to the set of axonal termination coordinates (Fig. 1D). Details of the numerical algorithm are described in Methods section. Such an optimization can be performed for any distribution of chemical markers (Ephs and ephrins). We next examine maps in conditions when wild-type distribution of chemical markers is altered by genetic manipulations.

When one of the chemical markers (ephrin-A) is completely eliminated in our model (Fig. 2) the map is no longer continuous. Instead, the map consists of locally smooth domains separated by linear discontinuities, reminiscent of fractures observed in the maps of preferred orientation in visual cortex. In contrast to orientation fractures linear singularities observed here have differing displacement of the retinal coordinate associated with them, as shown in Fig. 2D. Two neighboring domains in SC may be from extreme distal or proximal positions in retina.

The emergence of smooth domains is induced in our model by the activity-dependent refinement of the map, which leads to attraction between axonal terminals with close retinal coordinates. Correlated activity provides local refinement of the map yielding ordered local domains. However, since activity-dependent attraction has a finite range in our model it cannot induce global sorting of axons. In the wild-type conditions such global ordering is provided by chemical cues (ephrins). In the absence of chemical cues (Fig. 2) nothing enforces global ordering, resulting in arrangement of domains, determined by random factors (Fig. 2E).

It is tempting to compare our result with the observations in homozygous ephrin-A2/A5 -/- knockout mice[11], in which almost all chemical labels acting along TN axis are eliminated. Focal DiI injections in retina always resulted in labeling of discrete puncta in SC in these mutants. In our model, tracing terminals of RGC axons originating from proximal retinal positions results in labeling of focal clusters in SC (Fig. 2) in agreement with experimental findings of Ref. [11]. We observe 1-5 disconnected clusters, depending on the size of the traced axonal pool. In view of granular map structure in Fig. 2D this suggest the existence of granules of smooth mapping separated by linear position fractures in mutant ephrin-A2/A5-/- mice.

Similarly to Ref. [11] mapping in ephrin-A2/A5 -/- conditions displays significant errors in the perpendicular DV/ML direction. These mapping errors are *not* induced by interactions between ephrin-A ligands and EphB receptors, since such interactions are not present in our model [Model section, equation (2)]. Instead, DV topography appears to lack activity-dependent refinement pertinent to the wild-type conditions, since interactions between neighboring domains are negligible in fractured maps. This is because axons in neighboring domains come from distal points in TN direction in retina, and their activity-dependent correlations are negligible. Consequently, although topography is precise inside domains, their relative arrangement is not affected by activity-dependent refinement. The remaining crude DV topography is established due to the presence of EphB/ephrin-B chemical markers in ephrin-A knockouts.

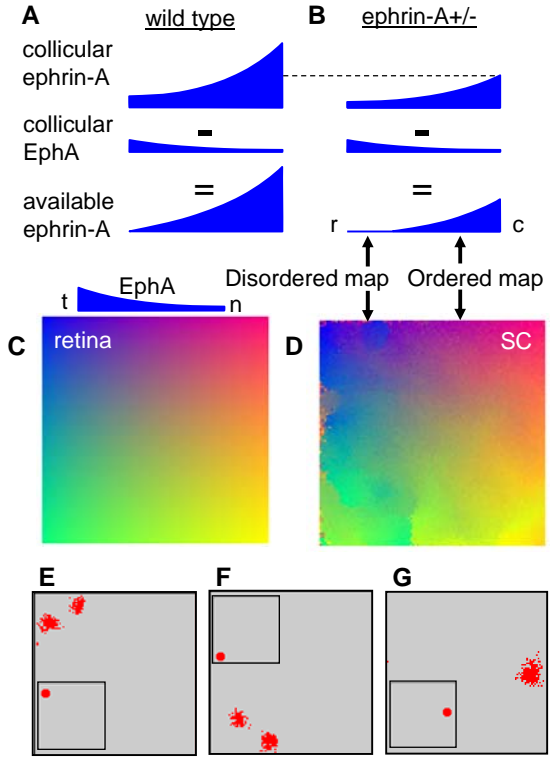

**Figure 3** Mapping in ephrin-A2/A5 heterozygous knockouts is intermediate between wild-type (Fig. 1) and homozygous cases (Fig.2). (**A**, **B**) Determination of the density of ligand available for signaling. The density of available ligand is reduced by binding to EphA receptor expressed in SC. (**B**) In case of ephrin-A +/- knockouts, when the density of expressed ligand is halved (dashed line), this may lead to almost complete masking in rostral SC. Retinocollicular map (**C**, **D**) becomes disordered in rostral SC, as in Fig. 2, while it is ordered in the caudal SC. Doubling of the map is evident in (**D**) from dual coloring in rostral SC. (**E-G**) In agreement with this, tracings of focally located RGC axons (insets) displays doubled/single clustered terminations in rostral/caudal SC respectively.

We next examine conditions when ephrin-A is partially reduced in SC[10, 11]. One way to understand these results is to include masking of ephrin-A level by EphA receptor expressed in SC. This implies that ephrin-A level available to convey positional signal is reduced by the presence of collicular EphA receptors (Fig. 3A,B). A similar phenomenon is observed in RGCs[11, 35, 36]. In the wild type mice masking leads to steepening of the ligand gradient (Fig. 3A)[11].

In ephrin-A2/A5 +/- mice the level of ligand is reduced by a factor of two compared to the wild type. This creates the potential for overmasking, which arises when the level of compensating receptor is higher than the level of ligand. This condition is similar to the loss of outgrowth sensitivity by temporal RGC axons with overexpressed/added ligand observed in striped assays[36, 37]. The same condition can be met in rostral SC, where the density of ligand is minimal (Fig. 3B). In case of overmasking, there is very little of ligand in rostral SC available for positional signaling (Fig. 3B), similarly to complete ephrin-A knockouts (Fig. 2D). The topographic map becomes hybrid: it is ordered in caudal SC similarly to the wild-type and disordered in rostral SC as in homozygous knockouts (Fig. 2D). Naturally, focal retinal injections lead to multiple puncta in rostral SC and single focal spots in caudal SC (Fig. 3E-G), as observed experimentally[11]. The overmasking phenomenon can also account for phenotypes observed in ephrin-A2-/- and ephrin-A5-/- knockout mice (Fig. 4, 5).

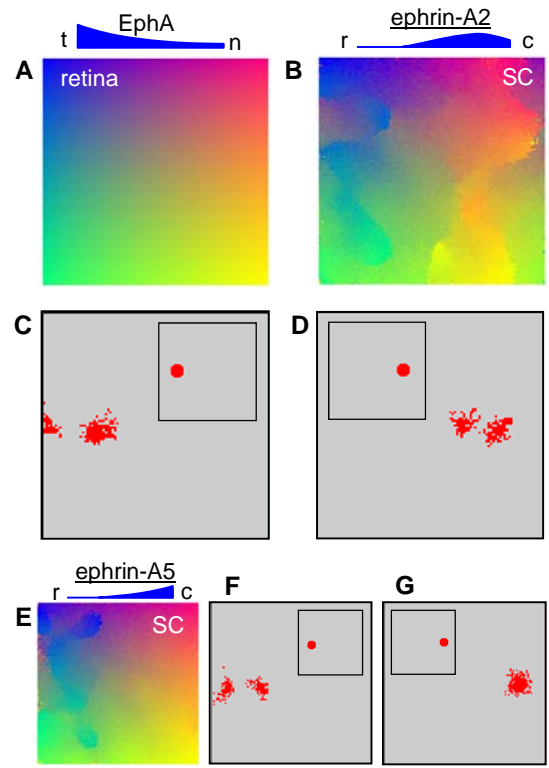

**Figure 4** Mapping in the conditions when a single ligand of the ephrin-A family (A5) is removed completely. (**A**, **B**) Complete map structure in a homozygous ephrin-A5 -/- conditions. The remaining ligand (ephrin-A2) is peaked in SC[10, 11]. Injections into both temporal (**C**) and nasal (**D**) retina lead to dual terminations in SC. (**E**) Complete map in a typical ephrin-A2 -/- knockout (ephrin-A5 is the remaining ligand). Injection into temporal retina (**F**) leads to two clustered terminations in rostral SC, while the nasal injection (**G**) results in a single focal group of points in SC. This is similar to what is observed experimentally[11].

We also examined an alternative to overmasking mechanism for generating disordered maps in temporal retina (rostral SC). This mechanism relies on a smaller gradient of ligand in rostral SC, which could lead to earlier emergence of disordered maps there in the knockout conditions. We found that this mechanism can be implemented in a narrow range of parameters and requires a significant inhomogeneity in the ligand gradient. This mechanism is thus less likely than overmasking.

In about 20% of heterozygous mutants and ephrin-A2/A5 individual knockouts (Fig. 3, 4) we also observe singularities of topographic map localized near a point (Fig. 5) as revealed by gradient maps (Fig. 5B). Several locations in retina are represented around such singularities, as shown by a retrograde injection in Fig. 5C and D, and by enhanced range colormap in Fig. 5B. In this respect such point singularities are similar to pinwheels in the maps of orientation preference in visual cortex[24-26]. The defects shown in Fig. 5 consist in removal of axons originating from retinal locations surrounded by the oval in Fig. 5C and projecting them elsewhere to a topographically inappropriate location. We did not observe point singularities in homozygous ephrin-A2/A5 knockouts.

## DISCUSSION

We present a quantitative model which can account for the experimental observations of retinocollicular mapping in ephrin-A knockout mice[10, 11]. These experiments usually involve tracing of a small group of axons from retina to SC. The use of computational techniques allows us to reconstruct complete map structure and to observe intriguing details of RGC axon arrangement in the absence of the chemical cues. The main finding here is that retinocollicular mapping, dominated by the activity-dependent mechanisms, lacks global ordering and contains domains of locally smooth mapping separated by linear singularities. These linear defects are similar to fractures in the orientation maps observed in higher brain areas, such as primate or feline primary visual cortex.

### The use of optimization principle

Development of sensory maps may be described by two methods. The first method involves specifying special rules followed by axons and dendrites during development, such as the rules of interaction with chemical cues[38], learning rules, etc.[39]. The second strategy implements neural development as optimization process. These two methods in many cases are found to be equivalent[27, 40]. Here we use the latter method for the following reason. Axons in developing nervous system are known to follow gradients of chemical labels[16] thus optimizing their affinity to targets. The problem of retinotectal development involves competition between axons for space and/or positive factors, such as neurotrophins, in the target [2, 8, 11, 18-21]. This system cannot be modeled by considering rules for single axons in isolation. Instead, we propose that the entire system of axons follows the gradient of cohesive energy functional (Fig. 1D) thus optimizing their total binding energy to the target. Treating formation of the map as a system allows accounting for possible tradeoffs between different parts of axonal ensemble due to competition. Thus, some axons may appear to be repelled by an attractive chemical label, because other axons are present in the population, which exhibit a stronger affinity to this same label[23].

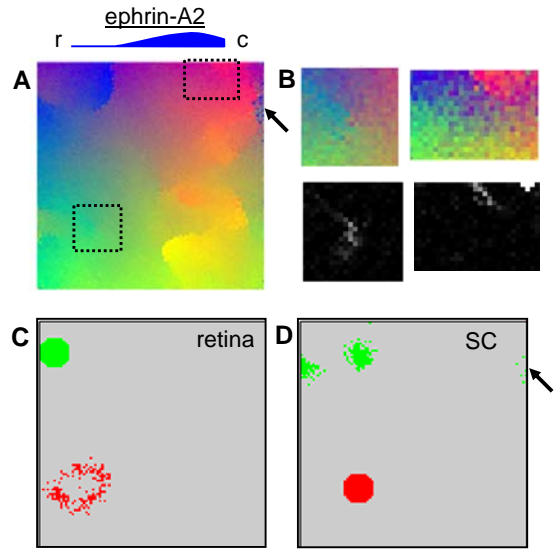

**Figure 5** Other mapping defects. (**A**) Ephrin-A5 -/- knockout conditions, similar to Fig 4B. The remaining ephrin-A2 peaks in caudal SC. Dotted boxes: regions containing point singularities enlarged in (**B**). Arrow shows regions filled with temporal axons (blue) in caudal SC[10]. (**B**) Regions boxed in (**A**) display two point singularities using enhanced colormap. Gradient maps (bottom) confirm that singularities are localized. (**C**, **D**) The anterograde injection into temporal retina (green) shows additional terminations in caudal SC (arrow). This is observed in Ref. [10] and in about 30% of our simulations. Retrograde injection into the center of point singularity (red) results in a ring in retina, implying that several positions are represented in the singularity, similarly to pinwheels in orientation maps.

### Additive interaction between correlated activity and chemoaffinity-dependent contributions

Our findings provide insights into the possible logic of interaction between chemical cues and activity-dependent mechanisms. Indeed, although activity block/reduction affects the precision of retinotectal/collicular mapping, rough topographic order is retained, presumably due to chemical cues[12, 14]. Thus, chemospecificity alone is sufficient for establishing crude connection specificity. This is accounted for by the plus sign between correlated activity-dependent and activity-independent terms in equation (1), which implies that either the former or the latter suffice. However, if chemical labels are eliminated, as in experiments with ephrin-A knockout mice, presumably, leaving the activity-dependent contribution intact, the resulting map appears to be random, which leads to an apparent contradiction with the additive logic of interaction between two factors. Here we resolve this apparent paradox. We suggest that the map is not entirely random in ephrin-A knockouts. Correlated activity-based mechanisms refine the map on the short scale, making the mapping topographically precise within the smooth domains. Thus, the additive logic of interaction between two factors does not contradict to the findings in ephrin-A deficient mice[10, 11]. Instead, chemospecificity and correlated activity may operate on different spatial scales: the former ensuring the global ordering, while the latter providing local refinement.

An alternative to our model uses multiplicative interactions between two factors[41]. Interestingly, such model finds its correlates in hippocampus, where induction of long-term potentiation is found to be Eph/ephrin-dependent[42].

**Mapping errors in the dorso-vertral direction**
Ref.[11] shows that removal of ephrin-A2/A5 ligands, which are thought to act as topographic labels along TN retinal axis, results in mapping errors in the perpendicular DV direction. Ref.[11] suggests that these DV mapping errors may follow from interactions between EphB and ephrin-A families of labels, which are known to exist[5, 29-31]. An additional support for this hypothesis comes from recent work[32], which shows that ephrin-A5, can bind and activate EphB2 receptor, expressed by RGC axons. Our results suggest an alternative explanation to the topographic errors in DV direction. Our model does not include interactions between labels responsible for mapping in the TN and DV directions [equation (2), Methods]. Yet, results for ephrin-A knockout conditions display significant errors along DV direction (Fig. 2D, 3E and F), as observed experimentally[11]. In our model these errors arise because activity-dependent refinement fails to act along DV direction, due to disruption of neighborhood relationships in ephrin-A knockouts (see above). Hence, the design of our model helps distinguish two hypotheses for DV errors.

**Experimental predictions**
Our findings can be verified explicitly if the structure of topographic map in ephrin-A mutants is revealed by optical imaging of intrinsic signals[43, 44]. An alternative implicit verification of the presence of smooth domains in SC would result from anterograde tracing of two or more groups of neighboring RGC axons by injecting several markers, such as DiI and DiAsp[9], into the same retina. According to our results these injections should cluster into closely positioned groups in SC (Fig. 5) if local topography is spared in ephrin-A mutants. Such experiments should tell if additive interactions between disparate factors are indeed involved in the formation of topographic maps.

**METHODS**

**Chemospecificity model.** We consider a square N by N (N=100) array of RGC axons projecting to an N by N array of termination sites in SC. Any terminal site can be occupied by one and only one RGC axon, which reflects competition. Each axon is characterized by the available (masked) levels of EphA/B receptors $R_{A/B}$ and each terminal site is described by an available concentration ephrin-A/B ligands $L_{A/B}$. The chemical cohesive energy is given by the mass action law

$$E_{chem} = \sum_i \alpha R_A(i) L_A(\vec{r}_i) - \beta R_B(i) L_B(\vec{r}_i). \quad (2)$$

Here $\alpha$ and $\beta$ are parameters describing the strength of contributions from binding of A and B receptors respectively; the sum is assumed over all axons numbered by index $i$ terminating at positions $\vec{r}_i$ in SC. Coefficients $\alpha=\beta=30$ are chosen to match experimentally observable RGC arborization diameter of about 40% of SC in the absence of correlated activity[14] (Fig. 1F).

To obtain available receptor/ligand levels we used the formulas $R_A = [R_{retA} - L_{retA}]_+$ and $L_A = [L_{collA} - R_{collA}]_+$, where the threshold-linear masking function is $[R]_+ = R$ for $R \geq 0$ and $[R]_+ = 0$ for $R < 0$. These formulas account for masking of retinal receptor by retinal ligand[11] and collicular ligand by collicular receptor. The masking function yields i.e. no available receptor when the level of ligand is larger. In the wild-type conditions the receptor expression levels by the axons is $R_{retA}(x) = \exp(-x/N)$, where $x = 1..N$ is the TN coordinate in retina. The total concentration of retinal ligand is $L_{retA}(x) = \exp(x/N - 2)$. The concentration of ligand/receptor in SC is $L_{collA}(x) = \exp(x/N - 1)$ and $R_{retA}(x) = \exp(-1 - x/N)$ respectively. The concentrations of signaling molecules $R_{retA}(x)$ and $L_{collA}$ are chosen from Refs.[18, 27], while the masking levels $L_{retA}(x)$ and $R_{collA}$ are selected not to exceed at any point the signaling concentrations.

To model maps in ephrin-A2/A5 knockouts we represent the net ligand level as a sum of two contributions: $L_{collA} = L_{collA2} + L_{collA5}$, where $L_{collA5} = a \exp(bx/N - 3) + c$. The coefficients $a, b, c = 0.66, 3, 0.15$ are chosen so that $L_{collA2}(x)$ is about the same in rostral and caudal parts (x=1 and N respectively)[10] and peaks at about x=0.7N[Ref.[10, 11]]. The masking level of ligands in retina is given by the same expressions, divided by a factor of $\exp(1)$ to prevent overmasking in the wild type conditions. We found that inclusion of receptor masking in retina does not affect our conclusions significantly. To model ephrin-A2/A5 knockouts the individual expression levels of these ligands are multiplied by a factor of ½ and 0 in heterozygous and homozygous conditions respectively in both retina and SC. The expression levels of EphB in retina and ephrinB in SC are assumed to be $R_B(y) = L_B(y) = \exp(y/N - 1)$, where $y$ is the DV/LM coordinate. No masking in the DV/LM direction is assumed for simplicity. The concentrations of available chemical labels are shown in the figures, when appropriate.

**Activity-dependent contribution.** Here we show that contribution to cohesive energy between the system of RGC axons and their targets due to correlated activity is described by pair-wise attraction in the form

$$E_{act} = -\frac{\gamma}{2} \sum_{ij} C_{ij} U(\vec{r}_i - \vec{r}_j). \qquad (3)$$

Here $\gamma$ describes the overall strength of this contribution, $C_{ij}$ is cross-correlation of electric activity between two cells, $U$ is the form-factor of attraction. In this study we adopt $C_{ij} = \exp(-r/R)$, where $r$ is retinal distance between axons $i$ and $j$, $R = 0.11N$ as obtained from the fit to cross-correlation due to retinal waves measured in Ref. [14]. $U(r) = \exp(-r^2/2d^2)$, $d = 3$, $\gamma = 0.25$. We have verified that our conclusions do not depend on the particular choice of parameters in a broad range.

Total retinal input at point $\vec{r}$ is determined by retino-collicular synaptic weights $w_i(\vec{r})$ and axonal activity $a_i$ through $I(\vec{r}) = \sum_i w_i(\vec{r}) a_i$. Activity of a collicular cell and point $\vec{r}$ is $A(\vec{r}) = \sum_{\vec{r}'} D(\vec{r} - \vec{r}') I(\vec{r}') = \sum_j v_j(\vec{r}) a_j$, where $D$ is dendritic form-factor, $v_j(\vec{r}) = \sum_{\vec{r}'} D(\vec{r} - \vec{r}') w_j(\vec{r}')$ is the net weight from RGC axon $j$ to the collicular cell. According to Hebbian rule the change in this weight is proportional to correlator between activities in this cell and the presynaptic cell (fire together, wire together)

$$\Delta v_i(\vec{r}) = \gamma \overline{A(\vec{r}) a_i} = -\partial E_{act} / \partial v_i(\vec{r}). \qquad (4)$$

The last equation interprets the weight adjustment as a gradient descent in $E_{act}$. Solving equation (4) for $E_{act}$ we obtain

$$E_{act} = -(\gamma/2) \sum_{\vec{r}} \sum_{ij} \overline{a_i a_j} v_i(\vec{r}) v_j(\vec{r}), \qquad (5)$$

which leads to (3) assuming $w_i(\vec{r}) = \delta_{\vec{r}\vec{r}_i}$, i.e. discrete axons, $C_{ij} = \overline{a_i a_j}$, and

$$U(\vec{r} - \vec{r}') = \sum_{\vec{r}''} D(\vec{r} - \vec{r}'') D(\vec{r}'' - \vec{r}'). \qquad (6)$$

This implies that the form-factor of axonal interactions in (3) is given by overlap of dendritic arbors.

**The minimization procedure.** We implement an algorithm similar to Metropolis Monte-Carlo scheme[45]. The details of the procedure are described in Refs. [23, 46]. Briefly, we start from a random map. On each step of the algorithm we choose two not necessarily adjacent axons in SC randomly and exchange them with probability $p = 1/[1 + \exp(4\Delta E)]$, where $\Delta E$ is the change in cohesive energy due to exchange. The probability of exchange is thus higher if the cohesive energy is lowered $\Delta E < 0$, leading to stochastic minimization of $E$. This step is repeated $10^7$ times. This procedure implements the Gibbs distribution in $E$ as shown in Ref. [27].